\documentclass[a4paper,10pt]{article}
\usepackage{natbib}
\usepackage[intlimits]{amsmath}

\newtheorem{Ipotesi}{Hypotesis}
\newtheorem{Definizione}{Definition}

\def\diff{d}
\def\veps{  {\varepsilon} }

\def\={\stackrel{\mathrm{def}}{=}}
\def\Zj{\mathcal{Z}_j}
\def\Zk{\mathcal{Z}_k}
\def\Zjk{\mathcal{Z}_{jk}}
\def\ej{\veps_{j}^{(1)}}
\def\ek{\veps_{k}^{(2)}}
\def\eint{\veps_{jk}^{int}}
\def\ejk{\veps_{j}^{(1)}+\veps_{k}^{(2)}}
\def\nujkb{ {\bar\nu}_{jk}}
\def\nujk{  {\nu_{jk}} }
\def\nujb{ {\bar\nu}_{j}}
\def\nuj{  {\nu_{j}} }
\def\nukb{ {\bar\nu}_{k}}
\def\nuk{  {\nu_{k}} }
\def\thetau{{\theta_1}}
\def\thetad{{\theta_2}}
\def\thetaud{{\theta_{12}}}
\def\thetauf{{\beta_1}}
\def\thetadf{{\beta_2}}
\def\thetaudf{{\beta_{12}}}
\def\sigmau{{\sigma_1}}
\def\sigmad{{\sigma_2}}
\def\sigmaud{{\sigma_{12}}}
\def\sigmauf{{S_1}}
\def\sigmadf{{S_2}}
\def\sigmaudf{{S_{12}}}
\def\Aexp{ {<\bar A>_U} }

\title{On the definition of temperature using time--averages.}
\author{A. Carati\footnote{Universit\`a di Milano, Dipartimento di Matematica
                           Via Saldini 50,  20133 Milano (Italy) E-mail: 
			   {\tt carati@mat.unimi.it}
			  }
       }
\date{\today}

\begin{document}

\maketitle

\begin{center} ABSTRACT \end{center}
This paper is a natural continuation of a previous one by the author,
which was concerned with 
the foundations of statistical thermodynamics far from
equilibrium. One of the problems left open in that paper was the correct
definition of temperature. 
In the literature, temperature is in
general defined through the mean kinetic energy of the particles of a
given system. 
In this paper,  instead, temperature is defined \emph{\`a la} Carath\'eodory, 
the system being coupled to a heat bath, and  temperature being 
singled out as the ``right'' integrating factor of the exchanged
heat. As a byproduct,
the ``right'' expression for the entropy is also obtained.  In
particular, in the case of a $q$-distributions the entropy  turns
out  to be that of Tsallis, which we however show to be additive, 
at variance with what is usually maintained.

\vfill
\newpage

\section{Introduction}
The problem of defining temperature in 
non--equilibrium situations is a quite delicate one (see for example
ref.~\cite{cipriani}).  
For systems constituted of particles, it is
usually assumed that temperature should be defined, up to a
constant factor, as  the
mean kinetic energy. On the other hand,
one should take into account that
the notion of temperature originates from thermodynamics, where
the notion of ``the particles of the system'' has no sense at all. In
thermodynamics (see refs.~\cite{libro}, \cite{wannier}) temperature is
defined using both the second and the zeroth principle. On the one
hand, the second principle (e.g. in the Carath\'eodory formulation) 
insures that there exists
an integrating factor of the exchanged heat. On the other hand,
since actually there exist infinitely many such integrating factors,
temperature is singled out among them by 
the zeroth principle, namely by the requirement that if the
system is put in thermal contact with another one, at 
equilibrium the integrating factors of the two systems and that of the
compound one should have the same value.

Thus,  when the Gibbs distribution $\exp(-\beta H)/Z(\beta)$ is used
in statistical mechanics, 
it is first checked  that $\beta^{-1}$ is an integrating
factor of the exchanged heat, but its identification with 
temperature requires some more work. One has to put the
system in contact with another one and to assume that (at equilibrium) the total
system  too is described by a Gibbs distribution having as Hamiltonian the
sum of the Hamiltonians of the two components. From this, with
some further considerations, one then shows
that the temperature coincides with 
$\beta^{-1}$ (see for example \cite{gibbs}, chapter three.). 

The aim of the present paper is to implement the analogous procedure 
when the averaging is performed through time--averages rather than through 
phase--averages with Gibbs' measure. 

For what concerns the time--averages, we follow  paper \cite{mio}, the 
 general set--up and the main results of which are recalled in Section~2. 
In Section~3 it will be recalled how temperature is introduced in
the standard thermodynamic way; moreover  
the definition of  ``thermal equilibrium'' between two systems
in terms of time-averages will be introduced, and  
it will be shown that temperature actually exists. 
Section~4 is devoted to the study of
the thermal contact between a generic system and a thermal bath
(a notion that will also be defined), while the
thermodynamics in the case of the Tsallis $q$--distribution will be
developed in Section~5. Section~6 is devoted to clarifying another
point that was left open in paper \cite{mio}, namely the
identification of the quantity $\alpha$ introduced there, which is
shown to be strictly related to free energy. This result is also used
for discussing the identification of temperature in the case of the
Tsallis $q$--distribution.
Finally,  the conclusions follow in Section~7.

\section{Time--averages}

For a system with phase space $\mathcal{M}$,
suppose a sequence $\{x_n\}$, $x_n\in\mathcal{M}$, is given, depending
parametrically on its first element $x_0$. As a particular case, one
can think of the orbit generated by the iteration of a map, for example
the time--$\Delta t$ map induced by the flow of an autonomous
Hamiltonian system.
Suppose we are interested in computing time--averages of a dynamical
variable $A(x)$ (a real function on $\mathcal{M}$)
$$
\bar A (x_0) {\=} \frac 1N \sum_{n=1}^N A(x_n) \qquad \hbox{for}\quad
N\gg1 \ ,
$$
the number $N$ playing the role of the ``final'' time, thought of as a
fixed parameter.  One can divide the space $\mathcal{M}$ into a large
number $K$ of disjoint cells $\Zj$ (such that $\mathcal{M}=\cup \Zj$),
and one has then
$$
\bar A(x_0) \simeq \sum_{j=1}^K A_j \frac {n_j}N \ ,
$$
where $A_j$ is the value of $A$ at a point $x\in\Zj$, and $n_j$ is the
number of times the sequence $\{x_n\}$ visits $\Zj$.  It is clear that
$n_j$ depends on $x_0$ so that, if a certain probability distribution
is assigned for the initial data $x_0$, correspondingly $n_j$ turns
out to be a random variable with a certain distribution function,
which will depend both on the dynamics (i.e. the map) 
and on the distribution of the initial data. So one can
speak in general of the ``\emph{a priori probability} 
$P$ that the cell $\Zj$
will be visited a number of times $n_j = n$'':
\footnote{Notice
that in \cite{mio} reference was instead made to the corresponding
cumulative distribution function $F_j(n)=P(n_j\le n)$.}
\begin{equation}\label{eq1}
P(n_j = n)=f_j(n) \ .
\end{equation}
For the sake of simplicity of the exposition, in paper \cite{mio} the
following hypothesis was introduced:
\begin{Ipotesi}\label{ip1}
The quantities $n_j$ are independent random variables,
conditioned by $\sum n_j=N$.
\end{Ipotesi}
This however is not at all necessary, and the computations could have
been performed without it, as will be shown below.

From the fact that the occupation numbers $n_j$ are random
variables, there follows that also the time--average $\bar
A(x_0)$ itself is a random variable, so that it is meaningful to
consider its expectation. Denoting by $<\cdot>$ expectation with
respect to the a priori distribution, one has then
$$
<\bar A> = \frac 1N \sum_{j=1}^K A_j <n_j> \ . 
$$
Now, in statistical thermodynamics  one does not  deal directly
with the a priori probability, because it is generally assumed that 
the time--average of a certain macroscopic quantity, typically the
energy of the system,
has a given value, which should play
the role of an independent variable. So we consider the energy of the system,
which we denote by $\veps$, and its time--average $\bar \veps
=\sum_j \veps_j n_j/N$, and we impose on the numbers $n_1, \cdots, n_K$
the condition
$$
 \frac 1N \sum_{j=1}^K \veps_j n_j = U  = \mathrm{const} \ .
$$
The problem of computing the \emph{a posteriori expectation} $\Aexp$
of $\bar A$ given $U$, 
is  solved, in the thermodynamic limit (see reference \cite{mio}), by the relation
\begin{equation}\label{eq2}
<\bar A>_U = - \frac 1N \sum_j A_j\chi'_j\left( \frac{\theta\veps_j}{N} + 
               \alpha \right) \ .
\end{equation}
where prime denotes derivative, and the function $\chi_j(z)$ is
defined through the Laplace
transform of the probability distribution function (\ref{eq1}) by
\begin{equation}\label{eq3}
\exp(\chi_j(z))\=\sum_{n=0}^{+\infty} e^{-nz}  f_j (n) \ ,
\end{equation}
while the parameters $\theta$ and $\alpha$ are determined by the
equations
\begin{equation}\label{eq4}
\left\{  
      \begin{array}{lcr}
          U & = & - \frac 1N \sum_j \veps_j \chi'_j\Big(\frac
                       {\veps_j\theta}N +\alpha \Big)  \\
          N & = & -   \sum_j \chi'_j\Big( \frac
                       {\veps_j\theta}N + \alpha \Big)  \ .  
      \end{array}
\right.
\end{equation}
In terms of the quantities 
\begin{equation} \label{eq4.5}
\nujb \= -\chi'_j \Big( \frac
{\veps_j\theta}N  + \alpha \Big) \ ,
\end{equation}
relations (\ref{eq2}) and (\ref{eq4}) take the form
\begin{equation}\label{eq5}
<\bar A>_U =\frac 1N \sum_j A_j\nujb \ , \quad
 U =\frac 1N \sum_j \veps_j \nujb \ , \quad
 N = \sum_j \nujb \ , 
\end{equation}
and this shows that $\nujb$ can be interpreted as the mean occupation
number of cell $\Zj$. 

In particular if the process of occupation of any cell is a
Poisson one, i.e. if the successive visits of a given cell 
are independent events, then one finds
\begin{equation}\label{eq4.4}
   \chi_j(z)= p\exp(-z)-p \ ,
\end{equation}
with a parameter $p>0$. 
In such a case one easily shows that the system follows a Gibbs
statistics. In fact the mean occupation numbers are  easily
calculated from (\ref{eq4.5}), and turn out to be given by
\begin{equation}\label{eq4.6}
   \nukb = N \frac  {e^{-\theta\veps_k/N}}{Z(\theta)} \ , 
\end{equation}
where $e^{\alpha} \= Z(\theta)=\sum_k
e^{-\theta\veps_k/N}$  is the usual partition
function, so that relations (\ref{eq5}) become the usual canonical
ones.

Let us now recall how thermodynamics was formulated in
reference~\cite{mio} in terms of time--averages. First of all,
following Gibbs and Khinchin (see refs.~\cite{gibbs},\cite{khin}), one
defines the exchanged heat  $\delta Q$ in
terms of the work performed by the system. Indeed if one defines the
macroscopic work $\delta \mathcal{W}$ performed by the system as 
$\delta \mathcal{W}=<\partial_{\kappa} H >_U \diff \kappa $, i.e.
as the expectation of the microscopic work performed when a parameter 
$\kappa$ entering the Hamiltonian is changed, then, using the first principle
one defines the exchanged heat as $\delta Q \= \diff U
-\delta \mathcal{W}$. It is easily shown (see ref.~\cite{mio},
but also below) that one has 
\begin{equation} \label{eq6}
    \delta Q = \frac 1N\sum_{j} \veps_j\diff \nujb \ .
\end{equation}
This expression shows,  recalling (\ref{eq4.5}), that it is convenient 
to introduce as an independent variable, in place of $z$,  the quantity
$\nuj=-\chi'_j(z)$,  and this naturally leads to introducing
in place of $\chi_j$ its Legendre transform $h_j$, defined as  usual by
$$
   h_j(\nuj)=\big( \nuj z+\chi_j(z)\big)\big|_{\nuj =-\chi'(z)} \ .
$$
Notice that, while $\bar \nuj$ has the meaning of a mean occupation
number (conditioned on $U$), the quantity $\nuj$ just plays the role
of a parameter, in the same sense as $z$ does in (\ref{eq3}). 
In particular, the
quantities  $\nuj$ do not need satisfy any condition related to
normalization, or the fixing of an energy value. 
Now, from the Legendre duality, one has
$$ 
\nujb    = -\chi'_j\left(\frac {\theta\veps_j}{N} + \alpha\right) 
\qquad \Longleftrightarrow \qquad
 \frac {\theta\veps_j}{N} + \alpha = h'_j(\nujb) \ ,
$$
so that, expressing $\veps_j$ in terms of $h'_j$ and using $\sum \diff
\nujb = 0$,  relation (\ref{eq6}) 
takes the form 
\begin{equation} \label{eq7}
    \delta Q = \frac {1}{\theta}\sum_{j}h'_j(\nujb) \diff
               \nujb =
               \frac {N}{\theta}\diff\left(\frac 1N \sum_{j}h_j(\nujb)\right) 
 \ .
\end{equation}
This shows that the exchanged heat always admits an integrating
factor. 

The problem left open in reference~\cite{mio} is that there
actually exist infinitely many integrating factors, so that a further
requirement is needed in order 
to single out which one should be identified with
the inverse absolute temperature.
The aim of this paper is to show that,  under
suitable hypotheses, $\theta/N$ indeed \emph{is} 
the inverse temperature, and 
consequently $h=\sum_{j}h_j(\nujb)$ is the thermodynamic entropy.

\section{Thermal equilibrium.}

As a general fact
it is well known that, if the exchanged heat $\delta Q$ admits an
integrating factor $\theta$ (i.e. 
$\theta\delta Q=\diff \sigma$ for a certain $\sigma$), 
then any function of the form $\theta F(\sigma)$
will be an integrating factor too. The (inverse) temperature
$\beta$ is singled out by the requirement that, if two
systems are put in thermal contact, at equilibrium the values of the
integrating factors are the same. More precisely, consider 
systems $A$ and $B$, with their integrating factors $\theta_1$,
$\theta_2$ and related functions $\sigma_1$ and $\sigma_2$. Put
them in thermal contact to form  system $C$, with  integrating
factor $\theta_{12}$ and function $\sigma_{12}$. Then one can show (as
recalled in the Appendix) that  for each of the systems there exists
essentially a unique integrating factor, characterized by the property
that the values of the three functions
$\beta_1$, $\beta_2$, and $\beta_{12}$ are equal when
$A$ and $B$ are in mutual equilibrium. By definition,
the inverse of this integrating
factor is the (absolute) temperature of the system, and the 
corresponding
functions $S_1$, $S_2$, $S_{12}$ are the entropies. 
From the equality of the integrating factors there follows immediately 
that one also has
\begin{equation*}
   \diff S_{12} = \diff S_1 + \diff S_2 \ ,
\end{equation*}
i.e.,  as one usually says, entropy is additive.
We want to implement now this thermodynamic approach in order to
identify the temperature of our system. So we have to couple our
system to another one, define the notion of mutual equilibrium and show
that  three corresponding integrating factors can be found such that at
equilibrium their values are equal. 

Consider two systems with phase spaces $\mathcal{M}_1$ and $\mathcal{M}_2$, let
$\{\Zj\}$ and $\{\Zk\}$ be the corresponding partitions into cells, 
and  $\ej$ and $\ek$ the corresponding values of the energy.
When the systems are isolated, they are supposed to be described as in Section~2. 
When put in thermal contact, they form a compound system with phase space
$\mathcal{M}_{12}=\mathcal{M}_1 \times \mathcal{M}_2 $,  and $\{\Zjk\}\= \{
\Zj\times\Zk\}$ is a partition with  corresponding  energies
$\ej+\ek+\lambda \eint$; here, the term $\lambda \eint$ corresponds to
the  interaction due to the thermal contact, which should be thought of 
as small (one may assume $\lambda\ll 1$). 
We denote by  $n_{jk}$ the numbers of times an orbit  $\{
(x_n^{(1)},x_n^{(2)}) \}$ in $\mathcal{M}_{12}$ visits the cell $\Zjk$,
and correspondingly we denote by  $n_j$ and $n_k$ the number of times
the corresponding projections $\{x_n^{(1)}\}$, $\{x_n^{(2)}\}$ visit
the cell $\Zj$ and $\Zk$ respectively.  
One obviously has the relation
\footnote{For the sake of
notational simplicity, here and in the following we write $n_j$ in
place of $n_j^{(1)}$, and $n_k$ in place of  $n_k^{(2)}$, with the
understanding that index $j$ refers always to system ~1 while index $k$
refers to system~2. The same is understood for other objects such as
for example $\Zj$, $\Zk$.}
\begin{equation*}
  \begin{split} 
    n_j &= \sum_{k} n_{jk} \\
    n_k &= \sum_{j} n_{jk} \ .
 \end{split}
\end{equation*}

Concerning the a priori probability distribution for the 
occupation numbers, this will in general depend not only on the
distribution  of the initial data
in the space $\mathcal{M}_1 \times \mathcal{M}_2 $, but also  on $\lambda$
(i.e. on the dynamics). One should however take into account
that, if the two systems are at the same temperature, nothing
happens when they are put in contact, i.e. the probability
distribution for $n_j$ and $n_k$ will not change, or rather  will change
so little that the changes can be neglected.
This in turn  implies that the probability
distribution on the product phase space cannot be given in an
arbitrary way,  
but must have some relation to the case $\lambda=0$.
With this motivation in mind we give the following definition of  
mutual thermal equilibrium
\begin{Definizione} \label{def:1} 
Two systems are said to be in \emph{mutual
equilibrium} if their a priori probabilities $f_j^{(1)}(n)$ and  
$f_k^{(2)}(n)$ do not depend sensibly on $\lambda$ for $\lambda\simeq0$.
\end{Definizione}

It can be shown (see the next section) that 
the notion of thermal equilibrium implies that the mean energies
$U_1$ and $U_2$ of the two
systems cannot be given at will, but (having fixed all external
parameters) the value of the energy of any of the two systems 
fixes the value of the energy of the other one.
Thus,
in the plane $U_1$, $U_2$ there remains defined an \emph{equilibrium
curve} which determines the relation that the energies of two systems
have to satisfy when they are in mutual equilibrium. 
 
To show that there exist integrating factors which
have the same value for the three systems, one needs taking into
consideration
an apparently obvious relation among the exchanged heats, namely
$$
\delta Q_{12} = \delta Q_1 + \delta Q_2 \ ,
$$
where $\delta Q_{1}$,  $\delta Q_{2}$ and $\delta Q_{12}$ are the
heats exchanged  by system 1, system 2 and the compound system
respectively. This relation is actually far from trivial, because
the single terms $\delta Q_{1}$ and $\delta Q_{2}$ are not a priori
the same ones as one would have in the absence of a thermal contact
between the systems. The relation is however true when there is a mutual
equilibrium, because
in such a case one has
\begin{equation*}
 \begin{split}
   \delta Q_{12} &=\sum_{jk} (\ej+\ek) \diff \nujkb=
		   \sum_{j} \ej \sum_{k} \diff \nujkb
		  +\sum_{k} \ek \sum_{j} \diff \nujkb \\
                 &=\sum_{j} \ej \diff \nujb
		  +\sum_{k} \ek \diff \nukb 
                  =\delta Q_1 + \delta Q_2 \ ,
 \end{split}
\end{equation*}
where $\nujkb$ is the expectation of $n_{jk}$ conditioned by $\sum
n_{jk}=N$ and $\sum(\ejk)n_{jk}=U$. 
In the above expression, 
in virtue of Definition \ref{def:1} 
the quantities $\nujb$, $\nukb$ are essentially the 
same as in the uncoupled case $\lambda=0$; by the same token the
contribution $\lambda \eint$ to $\veps_{jk}$ was neglected.  
From the above expression for $\delta Q_{12}$ and relation (\ref{eq7}) 
there follows that there exist
functions $\thetau$, $\thetad$, $\thetaud$, and correspondingly
$\sigmau$, $\sigmad$, $\sigmaud$,  such that
\begin{equation}\label{eq8}
\frac 1\thetaud \diff\sigmaud=  \frac 1\thetau \diff\sigmau + 
       \frac 1\thetad \diff\sigmad  \ .
\end{equation}
From this it can be proved that there exist
three integrating factors (the inverse temperatures)
$\thetauf$, $\thetadf$, $\thetaudf$, uniquely defined apart from a
multiplicative constant, 
which have the same common value for the three systems, and 
correspondingly three functions (the thermodynamic entropies)
$\sigmauf$, $\sigmadf$, $\sigmaudf$ which are additive in the sense
that
\begin{equation*}
\diff\sigmaudf =  \diff\sigmauf + \diff\sigmadf  \ .
\end{equation*}
The proof is standard and is recalled in Appendix~A. 

There remains the problem that, for the coupling of two generic
systems, we are presently unable to find an explicit expression for
the entropy of the compound system. We are able however to do it when
one of the systems is a heat bath.
In the next section the coupling of a system with a heat
bath is considered, and it is shown 
how to compute the probability distribution of the
occupation numbers $n_{jk}$ and the
thermodynamic quantities of interest.

\section{A system in contact with a heat bath.} 
We want to compute the probability distribution of the occupation
numbers $n_{jk}$ for the cells $\Zjk$ of the compound system
$\mathcal{M}_1\times\mathcal{M}_2$, when the second system is a heat
bath, i.e.  follows a Gibbs distribution. For the sake of simplicity
we will limit ourselves to the case in which the probability
$f_j^{(1)}(n)$ for the occupation number of the cell $\Zj$ of the first
subsystem does not actually depend on $j$, i.e. one has
$f_j^{(1)}(n)=f^{(1)}(n)$.
Coherently, the corresponding Laplace transform (\ref{eq3}) will be denoted by
$\exp(\chi(z))$. 
From Definition~\ref{def:1} one
can limit oneself to consider the uncoupled case $\lambda=0$. As recalled
above, in the case of a system described by a Gibbs distribution the
visiting of the cells are independent events having a common
probability $p$ to happen.

Our aim is now to compute $P(\{n_{jk} \})$, namely
the probability of a given
set $\{n_{jk}\}$ of occupation numbers. The main difference with
respect to the case considered in ref.~\cite{mio}, is that now the
random  variables $n_{jk}$ cannot
be assumed to be independent, so that now
$P(\{n_{jk} \})$ is not factorized. One can 
proceed in the following way.
For a given set $\{n_{jk}\}$ let $l_j=\sum_k n_{jk}$ be the
corresponding number of visits of cell
$\Zj$ in the first system. As the visits of the cells of the second system are
independent events, the probability distribution conditioned by fixing $l_j$
will be multinomial, i.e.  will be giving by
$$
P\Big( \{n_{jk}\} |l_j=\sum_k n_{jk} \Big) =
    \frac {l_j!}{n_{j1}!\cdots n_{jK_2}!}p^{l_j} \ .
$$
On the other hand, 
if $f^{(1)}(l_j)$ is the probability that the cell $\Zj$ of the first
system is
visited $l_j$ times, as the occupation numbers
of the first system have been assumed to be independent (Hypothesis~\ref{ip1}),
one  finally has
\begin{equation}\label{eq9}
 P(\{n_{jk}\})= \prod_{j} \frac {l_j!p^{l_j}}{n_{j1}!\cdots n_{jK_2}!} 
                f^{(1)}(l_j) \ .
\end{equation}
In computing the conditional expectations, it will be seen in the next
pages that
an essential role  is played by the Laplace transform of distribution (\ref{eq9}).
A simple computation shows that
\begin{equation}\label{eq10}
    \sum_{\{ n_{jk} \} } e^{-\sum_{jk}n_{jk}z_{jk}} P(\{n_{jk}\}) =
         \exp\Big[\sum_j \chi\big(-\log(p\sum_k e^{-z_{jk}})\big)\Big] \ ,
\end{equation}
where by $\sum_{\{ n_{jk} \}  }$ we mean a sum over all
possible sets $\{n_{jk}\}$.
In fact, (\ref{eq10}) follows from the chain of identities:
\begin{equation*}
 \begin{split}
    \sum_{\{n_{jk}\}} &e^{-\sum_{jk}n_{jk}z_{jk}} P(\{n_{jk}\}) =
       \sum_{\{l_j\ge0\}} \sum_{\{\sum_k n_{jk}=l_j\}}\prod_{j} 
       \frac {l_j!p^{l_j} e^{-\sum_{jk}n_{jk}z_{jk}}}
       {n_{j1}!\cdots n_{jK_2}!}f(l_j) \\
    &= \prod_{j} \sum_{ l_j\ge0} f(l_j) \sum_{\sum_k n_{jk}=l_j}
        \frac{l_j!(pe^{-z_{j1}})^{n_{j1}}\cdots(pe^{-z_{jK_2}})^{n_{jK_2}}}
          {n_{j1}!\cdots n_{jK_2}!}  \\
    &= \prod_{j} \sum_{ l_j\ge0} f(l_j) \big(\sum_k pe^{-z_{jk}}\big)^{l_j} \\
    &= \exp\Big[\sum_j\chi\big(-\log(p\sum_k e^{-z_{jk}})\big)\Big] \ .
 \end{split}
\end{equation*}

One has now to compute the expectation of the time--average of a generic dynamical
variable $A$, conditioned by $\sum(\ejk) n_{jk}=U$ and $\sum n_{jk}=N$, 
namely the quantity
\begin{equation*}
 <\bar A>_U= \frac 1N
   \sideset{}{'}\sum_{\{n_{jk} \}} \sum_{jk}A_{jk} n_{jk}\, P(\{n_{jk}\}) \quad \Big/  
   \sideset{}{'}\sum_{\{n_{jk} \}} P(\{ n_{jk}\}) \ ,
\end{equation*}
where $\sum'$ denotes a sum over the possible sequences $\{n_{jk}\}$
constrained by $\sum n_{jk}=N$ and $\sum (\ejk) n_{jk}=U$.
This can be reduced to the computation of the ``generating
function'' 
\begin{equation}
Z(A,\mu)\= \sideset{}{'}\sum_{\{n_{jk} \}} \exp(-\mu\sum
                       A_{jk}n_{jk}) P(\{ n_{jk} \}) \ ,
\label{eq11}
\end{equation}
through  the relation
\begin{equation}\label{eq12}
<\bar A>_U= -\left. \frac 1N \frac {\partial\ }{\partial\mu} \log Z(A,\mu) 
        \right|_{\mu=0} \ .
\end{equation}

It turns out that, as in reference \cite{mio},   the asymptotic
expansion of the generating function $Z(A,\mu)$ is very simply
computed in the limit of very
``large'' systems (the ones of interest for thermodynamics), by using
the steepest descent method. This indeed is commonly  done in statistical
mechanics, following  Fowler and Darwin (see ref.~\cite{fowler}).
Such an expansion  will be performed below up
to the leading term, neglecting the remainder (an explicit expression
of which could however be given).  As the remainder
depends both on the form of the energy of the total system, and on the
function $\chi(z)$ (i.e. on the probability distribution)
characterizing the first system, the
validity of the procedure should be checked for any particular
system. In paper \cite{mio} it was shown that such a procedure
is  indeed correct, for example, for  systems described by the Gibbs measure.
So, we suppose that our systems too are well described by the leading
order term of the asymptotic expansion, 
and we presently show  how the expansion is actually performed. 

This goes as follows. The expression (\ref{eq11}) can be rewritten as
\begin{equation*}
 \begin{split}
   Z(A,\mu) &= \sum_{\{  n_{jk}\ge 0 \}} \delta\Big( \frac 1N
             \sum(\ejk)n_{jk} - U \Big) \delta\Big( \sum n_{jk} - 
             N \Big)  \\
            &~\qquad 
                \exp (-\mu\sum A_{jk} n_{jk}) P \Big( \{ n_{jk}
             \} \Big) \\
            &= \lim_{L\to+\infty} \iint_{-L}^{L} \diff
               \kappa_1\diff\kappa_2 \exp(-i\kappa_1 U - i\kappa_2N) \\
            &~\qquad 
               \sum_{\{  n_{jk}\ge 0 \}} \exp (- \sum n_{jk}( \mu
               A_{jk} +\frac {i\kappa_1}N(\ejk) + i\kappa_2) P \Big( \{ n_{jk}
             \} \Big) \\
            &= \lim_{L\to+\infty} \iint_{-L}^{L} \diff
               \kappa_1\diff\kappa_2 \exp(-i\kappa_1 U - i\kappa_2 N) \\
            &~
               \exp\bigg[ \sum_j \chi\Big(-\log\big(p\sum e^{\mu
               A_{jk} + i\kappa_1(\ejk)/N + i\kappa_2 }\big)\Big)\bigg] \ ,
 \end{split}
\end{equation*}
where in the second line the familiar representation of the
Dirac delta function $\delta(x)= \int \diff \kappa \exp(i\kappa x)$
was used,
while in the third line use was made of formula (\ref{eq10}) 
for the Laplace transform of the probability $P(\{ n_{jk} \})$.
The (double) integral in the last line can be evaluated using the
steepest descent method, and to leading order one finds 
\begin{equation*}
  \begin{split}
  \log Z(A,\mu) &= - \frac {\theta}N U -\alpha N + \\
  &~\qquad
       \sum_j \chi\Big(-\log\big(p\sum e^{\mu
               A_{jk} + \theta(\ejk)/N + \alpha}\big)\Big) \ ,
 \end{split}
\end{equation*}
where $\theta$ and $\alpha$ are the  solution of the system
\begin{equation*}
 \begin{split}
   U &= - \frac 1N \sum_{jk} (\ejk) \chi' \left(-\log\Big(p\sum_k e^{\mu
               A_{jk} + \theta(\ejk)/N + \alpha}\Big)\right) \cdot \\
           &~\qquad\qquad \frac 
               {\exp\Big( \mu A_{jk} +\frac {\theta}N(\ejk) \Big)}
               {\sum_k \exp\Big( \mu A_{jk}+\frac {\theta}N(\ejk) \Big)} \\
   N &= - \sum_{jk} \chi' \left(-\log\Big(p\sum_k e^{\mu
               A_{jk} + \theta(\ejk)N + \alpha} \Big) \right) \cdot \\
           &~\qquad\qquad  \frac 
               {\exp\Big( \mu A_{jk} +\frac {\theta}N(\ejk) \Big)}
               {\sum_k \exp\Big( \mu A_{jk}+\frac {\theta}N(\ejk) \Big)} \ .
 \end{split}
\end{equation*}
Now, taking the derivative of $\log Z(A,\mu)$ and putting $\mu=0$,
after some simple algebra one finds 
\begin{equation}\label{eq13}
\Aexp = - \frac 1N \sum_{jk} A_{jk} \chi'\left( \frac {\theta}N \ej +
         \alpha +\log(pZ_2)\right) \frac
         {e^{-\theta\ek/N}}{Z_2(\theta)} \ ,
\end{equation}
where we have defined $Z_2(\theta)\=\sum \exp(-\theta\ek/N)$, whereas
the constants $\theta$ and $\alpha$ are the solution of the previous system
with $\mu=0$, i.e. are solution of
\begin{equation}\label{eq14}
 \begin{split}
   U &= - \frac 1N \sum_{jk} (\ejk) \chi' (\frac {\theta}N\ej + \alpha 
               + \log pZ_2) \frac {e^{-\theta\ek/N}}{Z_2(\theta)} \\
   N &= - \sum_{jk} \chi' (\frac {\theta}N\ej + \alpha 
               + \log pZ_2) \frac {e^{-\theta\ek/N}}{Z_2(\theta)} \ .
 \end{split}
\end{equation}
The formul\ae\ (\ref{eq13}) and (\ref{eq14}) solve the problem of
computing the conditional expectation of a dynamical variable of the
compound system, when the second one is a heat bath.

Now, it is very interesting to consider two limit cases: that in
which the observable $A$ depends only on the variables of the first
system (so that $A_{jk}=A_j$), and that in which $A$ depends only on
the variables of the second one ($A_{jk}=A_k$).
In the first case it is meaningful to consider a situation in which
the energy conditioning
is given not on the total energy of the compound system,
but on the energy $U_1$ of the first one. This 
essentially amounts to considering 
the first system as isolated from the second one. 
Now, if one computes the generating function with the conditioning 
$\sum \ej n_{jk}/N=U_1$, one finds
\begin{equation}\label{eq15}
<A>_{U_1}= - \frac 1N \sum_j A_j \chi'( \frac {\theta_1}N \ej
+\alpha_1) \ ,
\end{equation}
with  $\theta_1$ and $\alpha_1$ solution of
\begin{equation}\label{eq16}
 \begin{split}
   U_1 &= - \frac 1N \sum_{j} \ej \chi' (\frac {\theta_1}N\ej + \alpha_1) \\
   N   &= - \sum_{j} \chi' (\frac {\theta_1}N\ej + \alpha_1) \ .
 \end{split}
\end{equation}
These relations are the same as (\ref{eq2}) and (\ref{eq4}) of
Section~2, as it should be.

Instead, if the dynamical variable $A$ depends only on the variables
of the heat bath, and the conditioning is done  only on the energy $U_2$ of the
latter, one finds
\begin{equation}\label{eq17}
<A>_{U_2}= - \frac 1N \sum_k A_k \frac {\exp(-\theta_2\ek/N)}{Z_2(\theta_2)}
\ ,
\end{equation}
where $Z_2$ is the canonical partition function defined above, 
while   $\theta_2$ is defined by
\begin{equation}\label{eq18}
    \sum_{k} \ek \frac {\exp(-\theta_2\ek/N)}
           {Z_2(\theta_2)}=U_2 \ .
\end{equation}
These are the standard Gibbs relations.

It is interesting to compare these results with the computation of the
mean energy of any of the two systems, when \emph{the total energy $U$} is
fixed. One finds (using in the second line the definition of $Z_2$)
\begin{equation}
  \begin{split}
     <\bar \veps^{(1)}>_U & = 
          - \frac 1N \sum_{jk} \ej \chi'( \frac {\theta}N \ej +
         \alpha +\log pZ_2) \frac {e^{-\theta\ek/N}}{Z_2(\theta)} \\
     &=    - \frac 1N \sum_{j} \ej \chi'( \frac {\theta}N \ej +
         \alpha +\log pZ_2) \sum_k  \frac
    {e^{-\theta\ek/N}}{Z_2(\theta)} \\
    &= - \frac 1N \sum_{j} \ej \chi'( \frac {\theta}N \ej +
         \alpha +\log pZ_2) \ ,
 \end{split}
\end{equation}
which is the same as (\ref{eq16}) with $U_1=U_1(\theta) \=
<\bar \veps^{(1)}>_U $. For what concerns $U_2$ one finds instead 
(using now, in the second line, the second of (\ref{eq16}))
\begin{equation}
  \begin{split}
     <\bar \veps^{(2)}>_U &= 
          - \frac 1N \sum_{jk} \ek \chi'( \frac {\theta}N \ej +
         \alpha +\log pZ_2) \frac {e^{-\theta\ek/N}}{Z_2(\theta)} \\
     &=    - \frac 1N \sum_{j} \chi'( \frac {\theta}N \ej +
         \alpha +\log pZ_2) \sum_k \ek \frac
    {e^{-\theta\ek/N}}{Z_2(\theta)} \\
    &= \sum_k \ek \frac  {e^{-\theta\ek/N}}{Z_2(\theta)} \ ,
 \end{split}
\end{equation}
namely again the same as (\ref{eq18}), with $U_2=U_2(\theta) \=
<\bar \veps^{(2)}>_U $. 
These computations show first of all that  
\begin{equation}\label{eq20}
      U(\theta)=U_1(\theta) + U_2(\theta) \ ,
\end{equation}
but also, as mentioned in Section~3, that the equilibrium energies $U_1$ and
$U_2$ lie on a curve, i.e. the curve
$\left(U_1(\theta),U_2(\theta)\right)$ parametrized by $\theta$.

\section{The thermodynamics.}

We turn now to formul\ae\  (\ref{eq13}) and (\ref{eq14}), in order to
 write them in a more transparent way. In fact,
defining  in perfect analogy with (\ref{eq4.5}) the mean occupation
numbers $\nujkb$ by
\begin{equation}\label{eq21}
\nujkb \= -\chi'\left( \frac {\theta}N \ej +\alpha+\log pZ_2\right)
       \frac {\exp(-\theta\ek/N )}{ Z_2(\theta)} \ ,
\end{equation}
such formul\ae\ take the form
$$ 
\Aexp = \frac 1N \sum_{jk} A_{jk} \nujkb \ , \qquad
U = \frac 1N \sum_{jk} (\ejk) \nujkb \ , \qquad
N = \sum_{jk} \nujkb \ .
$$
As in the case of (\ref{eq4.5}), one has now
$$
\nujkb = -\left. \frac {\partial\quad}{\partial z_{jk}} \sum_{l'} \chi \left(
             -\log p\sum_l \exp(-z_{l'l}) \right) \right|_{ z_{l'l}=
         (\theta/N) ( \veps^{(1)}_{l'}+\veps^{(2)}_{l} ) + \alpha } \ .
$$

One can then introduce the Legendre transform
$h^{(12)}(\nu_{11},\cdots)$ of the function $\sum_{l'}\chi(\cdots)$ occurring
above, by
\begin{equation}\label{eq21.5}
  h^{(12)}(\nu_{11},\cdots) =  \sum_{jk} z_{jk}\nujk + \sum_{l'}\chi \left(
                 -\log p\sum_l \exp(-z_{l'l}) \right) \ ,
\end{equation}
where as usual the dependence of $z_{jk}$ on $\nujk$ is obtained by solving
\begin{equation} \label{eq22}
  \nujk = \partial_{z_{jk}}\sum_{l'} 
           \chi \left(-\log p\sum_l \exp(-z_{l'l}) \right)  \ .
\end{equation}
Now, the Legendre duality  gives
$$
 \frac{\theta}N (\ejk) + \alpha = \frac {\partial\quad}{\partial
         \nujk} h^{(12)} ({\bar\nu}_{11},\cdots) \ ,
$$
so that for the exchanged heat $\delta Q_{12}$ one finds
$$
\delta Q_{12}=\frac 1N \sum_{jk} (\ejk)\diff\nujkb = 
      \frac N{\theta} \frac 1N \sum_{jk} 
      \left( \frac{\partial h^{(12)}}{\partial\nujk} -\alpha \right) \diff
      \nujkb =  \frac N{\theta} \diff h^{(12)} \ .             
$$
This shows that the exchanged heat does indeed have an integrating
factor. 

But there is more. 
Indeed, from the definitions (\ref{eq4.5}) and
(\ref{eq21}) of the quantities $\nujb$ and $\nujkb$ respectively
one checks that
$$
\nujb =\sum_k \nujkb \ ,
$$
so that
$$
\frac 1N \sum_{jk} \ej \diff\nujkb = 
\sum_j \ej \diff \nujb = \delta Q_1 =\frac N{\theta}
\diff h^{(1)} \ ,
$$
where the last equality comes from (\ref{eq7}). In the same way one
can check that
$$
\frac 1N \sum_{jk} \ek \diff\nujkb = 
\sum_k \ek \diff \nukb = \delta Q_2 =\frac N{\theta}
\diff h^{(2)} \ .
$$
Here $h^{(2)}$ is the standard Gibbs entropy $h^{(2)} = \sum
n_k\log n_k -  n_k\log p $, while  $\nukb$ is given by (\ref{eq4.6}).

Finally, we find that
$$
\frac N{\theta} \diff h^{(12)} = \delta Q_{12}= \delta Q_1 + \delta Q_2 =
 \frac N{\theta} \diff h^{(1)} + \frac N{\theta} \diff h^{(2)} \ .
$$
This shows that $\theta/N$ is indeed the absolute temperature, while
the thermodynamic entropies can be identified\footnote{ Notice that in
paper \cite{mio} the thermodynamic entropy was denoted by
$S^{\mathrm{th}}$, while the quantity $h$ was denoted by $S$.} as 
$$
S_{1}(U_1,\kappa)\= h^{(1)}(\nujb) \ , \quad
S_{2}(U_2,\kappa)\= h^{(1)}(\nukb) \ , \quad
S_{12}(U,\kappa)\= h^{(12)}(\nujkb) \ . 
$$ 
Moreover the thermodynamic entropies
are additive, in the sense that
$$
\diff S_{1} + \diff S_{2} = \diff S_{12} \ .
$$
 
\section{The case of the $q$--distribution.}

If system~1 is described by a Gibbs statistics, the function
$\chi$ is an exponential, and the usual formul\ae\ are recovered. In
fact in such a case one gets
$$
\sum_j \chi\Big( -\log p \big( \sum_k\frac {\theta}N (\ejk)+\alpha\big)\Big)
       = \sum_{jk} p\exp \Big( -\frac {\theta}N (\ejk)-\alpha \Big)\ .
$$ 
From this, the additivity of entropy easily follows using the formula
$$
h=\sum_{jk} \nujkb \log\nujkb \ , 
$$
and the property
$$
\nujkb=\frac 1N \nujb\nukb \ ,
$$
which follows from definition (\ref{eq21}).
In fact, one has
$$
\nujkb = p e^{-\alpha}e^{-\theta\ej/N}e^{-\theta\ek/N} \ .
$$
On the other hand, 
computing  $\alpha$ by the second of (\ref{eq14}), one finds 
$$
pe^{-\alpha}\sum_{jk}e^{-\theta\ej/N}e^{-\theta\ek/N} = 
         \frac {N}{Z_1(\theta)Z_2(\theta)} \ ,
$$
so that for $\nujkb$  one gets
$$
\nujkb = N \frac {e^{-\theta\ej/N}}{Z_1(\theta)} 
           \frac {e^{-\theta\ek/N}}{Z_2(\theta)} =
           \frac 1N \nujb\nukb \ ,
$$
the second equality following from the known expression (\ref{eq4.6})
which holds both for $\nujb$ and $\nukb$.

A comment is now  in order before discussing the expression of the
entropy in the case system~1 follows the Tsallis $q$--distribution.
The relation $ \nujkb=\frac 1N \nujb\nukb$,
which can be rewritten in the more expressive form 
\begin{equation}\label{eq23}
<n_{jk}>_{U(\theta)} =\frac 1N
<n_{j}>_{U_1(\theta)}<n_{k}>_{U_2(\theta)} \ ,  
\end{equation}
holds always true, as one can check from the definition~(\ref{eq21})
of $\nujkb$   and from the definitions~(\ref{eq4.5}) and 
(\ref{eq4.6}) of $\nujb$ and $\nukb$. But 
the important point is that a 
relation of this kind holds neither for the occupation numbers $n_{jk}$, $n_j$,
$n_k$, nor for the parameters  $\nujk$, $\nuj$, $\nuk$.
In particular, concerning the two functions  
$h^{(12)}(\nujk)$ and $h^{(1)}(\nuj) + h^{(2)}(\nuk)$
(the values of which obviously cannot be compared unless some relation
is assumed between the corresponding arguments), we will find that 
their values coincide when they are computed at equilibrium,
i.e. when their arguments satisfy relation (\ref{eq23}). 

If system~1 follows the Tsallis $q$--distribution, then one has
\begin{equation}\label{eq222}
  \chi(z)= p_1(1-(1-q)z)^{1/(1-q)}-p_1
\end{equation}
with a constant $p_1>0$, 
so that (from \ref{eq21}) one gets
\begin{equation}\label{eq24}
   \nujk = p_1\left( 1+(1-q) \log pZ_j \right)^{q/(1-q)} 
           \frac {e^{-z_{jk}}}{Z_j}
\end{equation}
with $Z_j=\sum_k e^{-z_{jk}}$. The aim is now to compute the
explicit expression of the function $h^{(12)}$, the entropy of the
compound system, and make a comparison with 
the sum $h^{(1)} + h^{(2)}$. 

First one has to express $z_{jk}$ as a function of
$\nujk$. To this end,  note that from (\ref{eq24}), taking the
logarithm of both sides, one gets
$$
z_{jk} = -\log \nujk - \log\left( \Big(1-(1-q) \log pZ_j
         \Big)^{q/(1-q)} \right)
         + \log p_1 - \log Z_j \ .
$$ 
In turn,
the expression of the function $Z_j$ in terms of $\nujk$ is obtained
from (\ref{eq24}) by summing over the index $k$, which gives
$$
\sum_k \nujk = p_1\left( 1 + (1-q) \log pZ_j\right)^{q/(1-q)} \ ,
$$
so that
$$
\log pZ_j = \frac 1{1-q} \left[ \Big(\frac {\sum_k \nujk}{p_1}  
            \Big)^{\frac {1-q}q} -1 \right]  \ .
$$
Inserting this relation into the expression for $z_{jk}$ one finds
$$
z_{jk} = -\log \nujk + \log \Big( \sum_k \nujk\Big) + 
          \frac 1{1-q} \left[ 1 - \Big( \frac {\sum_k \nujk}{p_1} 
          \Big)^{\frac {1-q}{q}} \right]
          + \log p \ .
$$

It is now immediate to perform the Legendre transform $h^{(12)}$ of 
the function 
$\sum_j \chi(..)= \sum_j p_1(1-(1-q)\log(pZ_j)/q)^{1/(1-q)}-p_1$, and one
obtains
\begin{equation*}
  \begin{split}
h^{(12)} = - &\sum_{jk} ( - \nujk \log \nujk + \nujk\log p ) +
              \sum_j \big( \sum_k \nujk \big) \log \big( \sum_k \nujk \big) + \\
             & \frac 1{1-q} \sum_{jk} \nujk - \frac {p_1q}{1-q}   
               \sum_j  \Big( \frac {\sum_k \nujk}{p_1}
              \Big)^{\frac 1q}   \ .
  \end{split}
\end{equation*}
This expression reduces at equilibrium, i.e for
$\nujkb=\nujb\,\nukb/N$, to the  simpler one 
\begin{equation*}
h^{(12)} = \sum_k ( -\nukb \log \nukb + \nukb\log p ) +
              p_1 \frac { q \sum_j  \Big( \frac {\nujb}{p_1}
              \Big)^{\frac 1q} - \sum_j \Big( \frac {\nujb}{p_1}
              \Big)}{1-q}  \ .
\end{equation*}
The first term coincides with the familiar expression of the Boltzmann
entropy of the heat bath, whereas the second one coincides
with the expression of the Tsallis
entropy for the first system (see refs.~\cite{mio},\cite{tsallis}), 
written however in terms of the mean occupation numbers, rather than
in terms of the escort probabilities. It is then apparent that
$$
\diff h^{(12)} = \diff h^{(1)} + \diff h^{(2)}  \ .
$$

\section{The meaning of the parameter $\alpha$.}
We now come back to another problem concerning the case of a single system, 
which
was left open in paper \cite{mio}, namely the identification of the
quantity $\alpha$ entering formula (\ref{eq2}). To this end we remark that,
according to formula (\ref{eq4.5}), the mean occupation number 
is given by
\begin{equation*} 
\nujb \= -\chi'_j \Big( \frac
{\veps_j\theta}N  + \alpha \Big) \ ,
\end{equation*}
whereas the parameter $\theta/N$ was shown to be the temperature of
the system. It is thus convenient to choose as
independent variables $\beta\=\theta/N$ and $N$ in place of $U$ and
$N$. In fact if we define 
$$
f(\beta) \= \frac 1N \sum_j \chi_j ( \beta \veps_j  + \alpha ) \ ,
$$
taking the derivative of both sides with respect to $\beta$ and using
(\ref{eq5}) one finds (we denote  
$\frac \partial {\partial\beta} = \partial_{\beta}$)
$$ 
\partial_\beta (- \alpha-f) = U \ .
$$
Now we recall that the free energy $F=U-\beta^{-1}S$ satisfies the relation 
$$
\partial_\beta (\beta F) = U \ ,
$$
so that one has, apart from an additive constant (possibly depending on the
external parameters entering the Hamiltonian), the important relation 
$$
\alpha = -\beta F -f \ .
$$

In particular, if the system follows a Gibbs statistics, $f$ is a
constant and one has 
$$
\alpha = -\beta F \ . 
$$
This coincides with the familiar relation $ \beta F = - \log Z$,
where $Z$ is  the canonical partition function,
because in such a case one has $\alpha = \log Z$ (see formula (\ref{eq4.6})).

If instead the system follows a Tsallis distribution of index $q$, then the
expression of $\chi(z)$ is given by (\ref{eq222}), so that 
$f$ is given by
\begin{equation*}
 \begin{split}
    f(\beta) &= \frac 1N \sum_j \Big[ p_1\big( 1 - (1-q) (\beta \veps_j  +\alpha)
                       \big)^{\frac 1{1-q}} - p_1 \big] \\
            &=\frac 1N \sum_j p_1\big( 1 - (1-q) (\beta \veps_j  +\alpha)
                       \big)\big( 1 - (1-q) (\beta \veps_j  +\alpha)
                       \big)^{\frac q{1-q}} -  1 \\
           &= 1-(1-q) (\beta U +\alpha) -1 = (q-1)(\beta U +\alpha)\ ,
 \end{split}
\end{equation*}
where, in the second line, use was made of the second and the third of
(\ref{eq5}).  We find in this way
\begin{equation*}
  \alpha = - \frac 1q \, \beta \big[ F - (1-q)U\big] \ ,
\end{equation*}
which gives the relation between $\alpha$ and the thermodynamic
functions $F$ and $U$. 

At this point, it is worth noting that usually the Tsallis distribution is
written in the form (see ref.~\cite{tsallis})
\begin{equation*}
   \nujb = C(\beta_q) \big( 1 - \beta_q \veps_j )^{\frac q{1-q}} \ ,
\end{equation*}
where the function $C(\beta_q)$ is a 
normalization factor, and $\beta_q$ is determined by somehow
fixing the mean energy. But in terms of our $\beta$, i.e. of inverse
temperature, one has
\begin{equation*}
   \beta_q =\frac {(1-q) \beta}{1 - (1-q)\alpha} \ .
\end{equation*}
This shows that  $\beta_q$ is \emph{not} the inverse temperature, 
but a complicated function of
it, which could be obtained by expressing $\alpha$ 
as a function of $\beta$.

\section{Conclusions.}

So we have shown, that the
parameter $\beta\=\theta/N$ is the inverse temperature, in the sense
that it has the same value for every system which is in thermal
equilibrium with a heat bath at (inverse) temperature $\beta$.  Such
an identification also enables one to find out the thermodynamic entropy $S$. 
In particular, in the case of the Tsallis $q$--distribution, the entropy 
just coincides with his $q$--entropy (as one could have imagined). 
A relevant point is however that in such a case, at variance with what is
usually maintained, the
entropy turns out to be additive, at least for what concerns
its differential, i. e. in the sense that one has
$\diff S_{12} = \diff S_1 + \diff S_2 $. Whether such a relation
can be integrated to
give $S_{12} = S_1 + S_2 $ is a non trivial point which we are 
unable to discuss at the moment. 
In the literature there is a long debate about this
point; in particular it is often pointed out that, if entropy is assumed to
be additive (for independent systems), then the Boltzmann--Gibbs expression
should follow (see ref.~\cite{tsallis2}). 
This is usually based on a uniqueness theorem of Khinchin 
(see ref.~\cite{khin2})
in the information theory framework, in which additivity plays a
key role. 

Without entering these very interesting questions, we only
want to point out that, in our definition (\ref{eq21.5}) of the entropy, 
the independent variables are  the parameters $\nuj$, which
are \emph{not} the probabilities of the occupation of
the cells in the space phase. 
So the functional dependence of the entropy
on the quantities $\nuj$ may be out of the reach of
Khinchin's theorem, in which the role of the independent variables is
played by the probabilities.
In any case, it is true that one ought to understand in a deeper way
the connection of the present approach 
with Kinchin's theorem, and more generally with information entropy. 


\appendix

\section*{Appendix: proof of the existence of temperature.}
In the following,  for the sake of completeness, the familiar
deduction of the existence of absolute temperature is recalled. 
The only difference with respect to the treatments of most
textbooks, is that we make use not of the concept of ``empirical
temperature'', but rather of the concept, recalled above, of the
``equilibrium curve'' in the plane $(U_1,U_2)$.

So let us assume  there exist $\theta_1$, $\sigma_1$ 
such that for the exchanged
heat $\delta Q_1$ of system~A one has $\delta Q_1=\theta_1 \diff
\sigma_1$, and analogously for system~B and the compound 
system~A$\cup$B, i.e. 
$\delta Q_2 = \theta_2 \diff \sigma_2$, $\delta Q_{12}=
\theta_{12}\diff \sigma_{12}$. From $\delta Q_{12}= \delta Q_{1} +
\delta Q_{2} $, one has   
\begin{equation}\label{eqa0}
    \theta_{12} \diff \sigma_{12} = \theta_1 \diff \sigma_1 + 
                                    \theta_2 \diff \sigma_2 \ ,
\end{equation}
i.e. 
\begin{equation*}
  \diff \sigma_{12} = \frac {\theta_1}{\theta_{12}}\diff \sigma_1 + 
                      \frac {\theta_2}{\theta_{12}}\diff \sigma_2 \ .
\end{equation*}
This shows that $\sigma_{12}$ depends only on the values of $\sigma_1$
and $\sigma_2$, and not on any other external parameters entering 
the energies of system~A or system~B, i.e one has
$$
\sigma_{12}=G(\sigma_1,\sigma_2) \ .
$$
So  there follows  that the ratio of $\theta_i$ by
$\theta_{12}$
is equal to the partial derivative $\partial_{\sigma_i} G$, i.e.  
one has 
\begin{equation}\label{eqa1}
 \theta_1= \frac {\partial G}{\partial\sigma_1} \theta_{12}\ ,   \quad
 \theta_2= \frac {\partial G}{\partial\sigma_2} \theta_{12}   \ .
\end{equation}

Use now the condition  that systems A and B are in mutual
equilibrium. This means that, if one fixes all the parameters but the
energies $U_1$, $U_2$ of the two systems, then they lie on a curve
$(U_1(\rho),U_2(\rho)$ in the plane $(U_1,U_2)$. For example, in the
case of two gases, one can fix the volumes and think of changing the
internal energies of the gases through an isocoric transformation,
i.e. by heating or cooling the gases. In our case, instead, one can think
of fixing the entropy and changing $U$ by adiabatic transformations. In other
terms, one uses as independent variables the entropies and the
internal energies (and, perhaps, some further parameters 
if the former are not sufficient to the complete thermodynamic 
description of the systems).

Now, taking the logarithmic derivative of the expressions
(\ref{eqa1}) with respect to $\rho$, i.e. the variable which 
parameterizes  the equilibrium curve, one finds
$$
\partial_{\rho} \log \theta_1 = \partial_{\rho} \log \theta_{12} =
\partial_{\rho} \log \theta_2 \ .
$$
In principle, here, $\partial_{\rho} \log \theta_1$
depends only on $\rho$ and on the other parameters of the first
system, while $\partial_{\rho} \log \theta_2$ 
depends only on $\rho$ and the parameters of the
second one. This implies that actually all the three expressions have to
be equal to a function depending only on $\rho$ (as can be seen,
for example, by the fact that the derivatives with respect to the
parameters other then $\rho$
vanish). One has thus
$$
\partial_{\rho} \log \theta_1 = \partial_{\rho} \log \theta_{12} =
\partial_{\rho} \log \theta_2=f(\rho) \ ,
$$
which on  integration gives
\begin{equation}\label{eqa2}
\theta_1 = e^{F(\rho)} \Sigma_1(\sigma_1) \, , \
\theta_2 = e^{F(\rho)} \Sigma_1(\sigma_2) \, , \ 
\theta_{12} = e^{F(\rho)} \Sigma_{12}(\sigma_{1},\sigma_{2})\ ,
\end{equation}
where $F(\rho)$ is a primitive of $f(\rho)$ and $\Sigma_i$ are
integration constants. One can wonder why
$\Sigma_i$ depends only on $\sigma_i$.  This
too  follows from (\ref{eqa1}), which shows that the ratio
$\theta_1/\theta_2$ depends only on $\sigma_1$, $\sigma_2$, and not
on any other external parameters needed to describe  the
system. 

The last step is to show  that $T^{-1}\=\exp(F(\rho))$ is the integrating
factor one is looking for. 
By construction, $T^{-1}$ has the same value for all the three systems. It
is uniquely defined apart from a multiplicative constant
depending on the choice of the primitive $F$. 
To show that it is  an
integrating factor, one has to consider
relation (\ref{eqa0}), which,  after simplifying $T$, reads
$$
 \Sigma_{12}\diff \sigma_{12} = \Sigma_1\diff \sigma_1 +
                                \Sigma_2\diff \sigma_2 \ .     
$$
Thus one can define the entropy of the two systems as $\diff S_1 =
\Sigma_1(\sigma_1) \diff \sigma_1 $ and $\diff S_2 = \Sigma_2(\sigma_1)
\diff \sigma_2 $ respectively. Moreover, as one has
$$
 \Sigma_{12}\diff \sigma_{12} = \diff S_1 + \diff S_2 = \diff (S_1 +S_2) \ ,
$$
one finds in the first place that  $\Sigma_{12}\diff \sigma_{12}$ too is a total
differential $\diff S_{12}$, and furthermore that the entropies are
additive in the sense that
$$
 \diff S_{12} = \diff S_1 + \diff S_2 \ .
$$
This also shows that $T^{-1}$ is an integrating factor, because   one has
$$
\delta Q_1 = T \diff S_1\, ,  \ 
\delta Q_2 = T \diff S_2\, ,  \ 
\delta Q_{12} = T \diff S_{12} \ .
$$

\end{document}